\documentclass{llncs}
\usepackage{llncsdoc}
\usepackage{graphicx}
\usepackage{amsmath}
\usepackage{array}
\usepackage{multirow}
\usepackage{amsmath}
\usepackage{mathtools}
\usepackage{algorithm,algorithmicx}
\usepackage[noend]{algpseudocode}
\usepackage{color}
\usepackage{blindtext}
\begin{document}

\title{Decompose-and-Integrate Learning for Multi-class Segmentation in Medical Images}

\author{Yizhe Zhang$^{1}$, Michael T. C. Ying$^{2}$,
Danny Z. Chen$^1$
} 

\institute{$^1$ Department of Computer Science and Engineering, University of Notre Dame, Notre Dame, IN 46556, USA\\
	$^2$ Department of Health Technology and Informatics, The Hong Kong Polytechnic University, Hung Hom, Hong Kong}

\maketitle

\begin{abstract}
 Segmentation maps of medical images annotated by medical experts contain rich spatial information.  In this paper, we propose to decompose annotation maps to learn disentangled and richer feature transforms for segmentation problems in medical images. Our new scheme consists of two main stages: {\it decompose} and {\it integrate}.  {\it Decompose}: by annotation map decomposition, the original segmentation problem is decomposed into multiple segmentation sub-problems; these new segmentation sub-problems are modeled by training multiple deep learning modules, each with its own set of feature transforms. {\it Integrate}: a procedure summarizes the solutions of the modules in the previous stage; a final solution is then formed for the original segmentation problem. Multiple ways of annotation map decomposition are presented and a new end-to-end trainable $K$-to-1 deep network framework is developed for implementing our proposed ``decompose-and-integrate'' learning scheme. {\color {black}In experiments, we demonstrate that our decompose-and-integrate segmentation 
 scheme, utilizing state-of-the-art fully convolutional networks (e.g., DenseVoxNet in 3D and CUMedNet in 2D), improves segmentation performance on multiple 3D and 2D datasets. Ablation study confirms the effectiveness of our proposed learning scheme for medical images.}
 

\end{abstract}

\section{Introduction}
Segmentation annotation maps are crucial for supervised training a deep learning based segmentation model. For segmentation annotation maps, besides the class label dimension, there are spatial dimensions that contain rich information of object size, shape, and between-object/between-class relations.

Previous work has proposed some methods for modifying annotation maps for training better deep learning based segmentation models. Directional map \cite{uhrig2016pixel} was proposed to generate additional training loss based on the relative positions of the pixels to the centers of their corresponding objects. Deep watershed transform \cite{bai2017deep} provided a similar approach that converts an annotation map to a watershed energy map to guide the training of a segmentation model. These efforts demonstrated that changing segmentation annotation maps to include additional information (e.g., relative position, stronger instance-level information) can help train better deep learning models for segmentation tasks.

In medical image segmentation, different classes of objects often have strong mutual locations and spatial correlations. Due to these correlations, learning representation and feature transform for one object class can often indicate the existence of some other object classes (possibly nearby). A conventional way of using multi-class annotation maps is to treat a full annotation map as a whole subject and use spatial cross-entropy loss function to compare it with the model's outputs in back propagation \cite{ronneberger2015u,chen2016deep,zheng2018new}. Due to spatial correlations among different object classes, directly using annotations of all object classes to train a deep network may cause a deep network not to be able to fully exploring its representation learning ability for every object class, especially for those classes with small sizes and unclear/confusing appearance. Furthermore, there may be multiple distinct structures/clusters under one class of objects, and each sub-class structure may better utilize a unique set of feature representations. In principle, we believe that modeling individual classes and sub-classes of structures or objects can encourage deep learning models to learn richer and more comprehensive feature transforms and data representations for segmentation problems.


 \begin{figure}[t]
  \centering
  \includegraphics[width=4.5in]{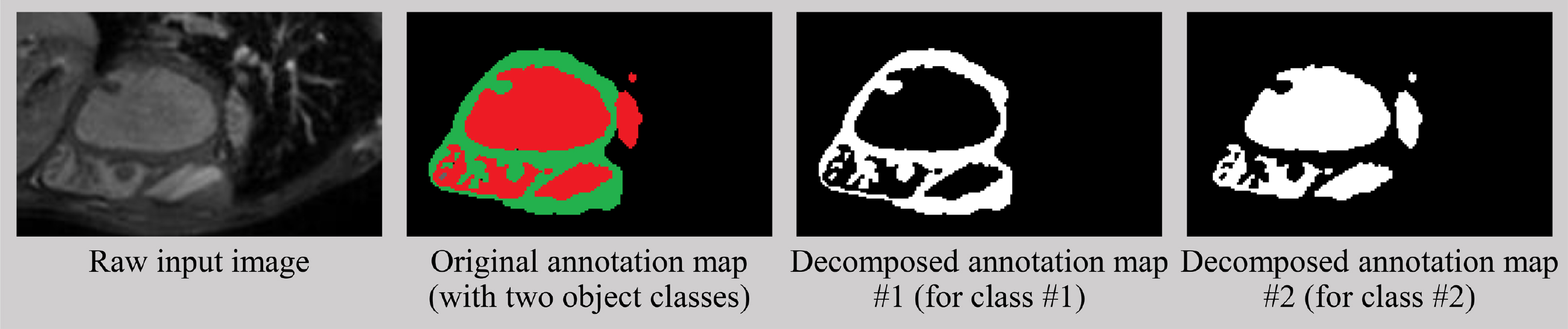}
  \caption{Segmentation annotation map decomposition based on object classes.}
\label{fig_decomposition_based_on_class}
\end{figure}

\begin{figure}[t]
  \centering
  \includegraphics[width=4.2in]{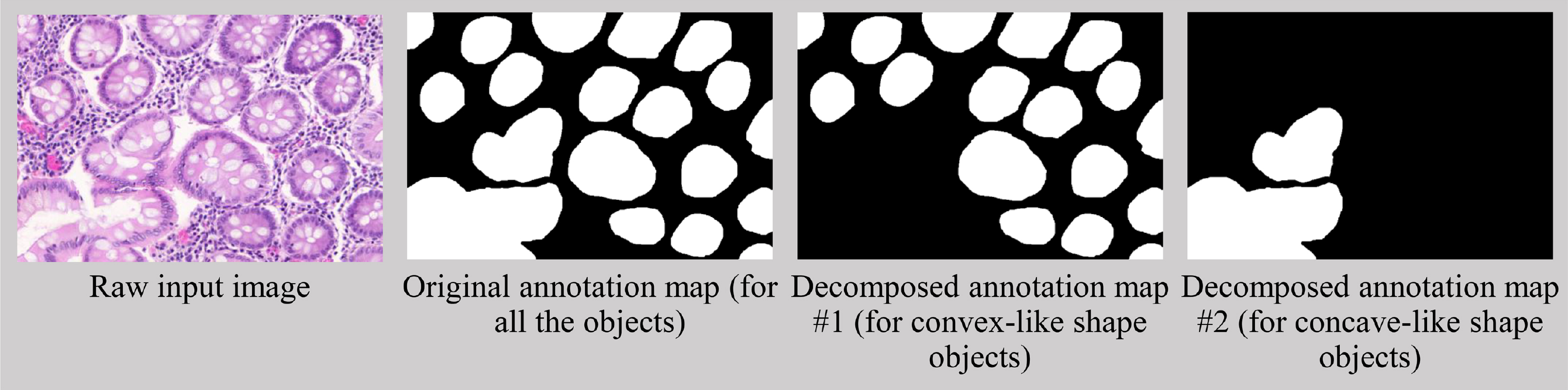}
  \caption{Segmentation annotation map decomposition based on object shape property.}
\label{fig_decomposition_based_on_shape}
\end{figure}

 In this paper, we propose to systematically decompose the original 
 annotation maps to encourage deep networks to learn richer and possibly more disentangled feature transforms and representations. Our new scheme consists of two main stages: {\it decompose} and {\it integrate}.  {\it Decompose}: by annotation map decomposition, the original segmentation problem is decomposed into multiple segmentation sub-problems (e.g., see Fig.~\ref{fig_decomposition_based_on_class}); these new segmentation sub-problems are modeled by training multiple deep learning modules, each with its own set of feature transforms. {\it Integrate}: a procedure summarizes the solutions of the modules in the previous stage; a final solution is then formed for the original segmentation problem.  This decompose-and-integrate scheme allows to explicitly enforce a deep learning model to learn representations for every object class. Besides, it can also be applied to learn feature transforms and representations for (human expert defined) meaningful sub-class data clusters and structures (see Fig.~\ref{fig_decomposition_based_on_shape}). 

In Section \ref{method_section}, we present different ways to decompose annotation maps for different scenarios, and develop a new $K$-to-1 deep network model for implementing our new  
 learning scheme. In Section \ref{sec:exp}, we evaluate our decompose-and-integrate learning scheme utilizing multiple state-of-the-art fully convolutional networks (FCNs) on three medical image segmentation datasets, and examine several proposed annotation decomposition (AD) methods.

\section{Decompose-and-Integrate Learning}\label{method_section}


Consider a $K$-class segmentation training dataset $\{(x_i,y_i)$, $i = 1,2, \ldots, h$\}, $x_i \in R^{m \times n}$ is a raw image, and $y_i \in \{1,2, \ldots, K\}^{m\times n}$ is a segmentation annotation map containing all the annotations of the $K$ classes of interest. Supervised learning for $K$-class segmentation tasks aims to learn a function $f \in F$ that transforms $x$ to $y$. Note that each $y_i$ can be denoted as $\{y_i^1,y_i^2, \ldots, y_i^K\}$, where $y_i^k\in \{0,1\}^{m\times n}$ is an annotation map for object class $k$, for $k = 1,2, \dots, K$.

For a segmentation problem with two foreground object classes $A$ and $B$, suppose modeling $p(y^B|x)$ for class $B$ is more difficult than modeling $p(y^A|x)$ for class $A$. This means that learning a robust latent representation $R^B(x)$ for class $B$ takes more computational effort (e.g., more training iterations/gradient descent effort) than learning a robust latent representation $R^A(x)$ for class $A$. 
Note that $R^B(x)$ and $R^A(x)$ are {\bf not} necessarily disjoint.
When $p(y^A)$ and $p(y^B)$ have moderate or high spatial correlations, using joint annotations of these two classes for training a deep learning model can lead to: 
(1) $p(y^A|x)$ is quite 
likely to be modeled using $R^A(x)$; (2) $p(y^B|x)$ would be modeled with help from $R^A(x)$, and not mainly by using $R^B(x)$; (3) $R^B(x)$ is not fully explored during model training, due to the ``help'' of the annotations from class $A$.
For a better representation and feature learning performance, such ``help'' is undesired. Besides multi-class segmentation scenarios, when an object class has distinct meaningful underlining sub-class structures/clusters, having a separate modeling for each individual structure/cluster enforces a deep network to learn more meaningful and useful data representations and feature transforms for such a class.

\subsection{Segmentation annotation map decomposition}\label{sec_AD}


\subsubsection{Based on object classes.}
For a $K$-class segmentation problem, we can decompose $y_i$ into $K$ binary annotation maps $y_i^k$, $k={1,2, \ldots, K}$. Algorithm \ref{alg:AD_class} gives the exact procedure. Fig.~\ref{fig_decomposition_based_on_class} shows an image illustration of the effect of this annotation decomposition (AD). In medical image segmentation problems, the number of object classes is usually small, and is much smaller than in natural scene images. 
A general guideline is that the decomposed segmentation maps and their associated extra computational costs should be under a manageable level. 
Table \ref{tabHVSMR1} shows that object-class based AD can effectively improve segmentation performance for segmentation problems with multiple foreground classes.

\vspace{-0.2in}
\begin{algorithm}
  \caption{Object-class based annotation decomposition}
  \label{alg:AD_class}
\footnotesize
  \begin{algorithmic}[1]
    \Function{AnnotationDecomposition1} {$y_i \in \{1,2, \ldots, K\}^{m\times n}$}
        \For{$k \gets 1 \textrm{ to } K$}
            \State{$y_i^k$ = a new array of size $m\times n$ with all 0;}
            \State {$y_i^k$ [where($y_i==k$)] $\gets 1$;}
        \EndFor
        \State \Return{$y_i^1, y_i^2, \dots ,y_i^K$}
    \EndFunction
  \end{algorithmic}
\end{algorithm}
\vspace{-0.4in}

\subsubsection{Based on object shapes.}
Annotation maps can also be decomposed based on different shape structures in the annotation maps. This type of decomposition can be applied to 2-class segmentation or even $K$-class segmentation for $K>2$. 


Shape information contains valuable cues for segmentation tasks. Decomposing annotation maps based on different object shapes can encourage a deep learning model to learn feature transforms that encode the raw images into different shape-guided representations. In histology image analysis, morphological features such as shape convexity play an important role in object detection, segmentation, and diagnosis. Thus, we propose to decompose segmentation annotation maps based on shape convexity of objects in the annotation maps. Specifically, two sub-segmentation maps are generated from an original segmentation annotation map, one containing convex-like shape objects and the other containing concave-like shape objects. {\color{black}This decomposition provides additional information that directly helps a learning model to perceive object information at a higher (object shape) level.} The detailed procedure and image illustration are provided in Algorithm \ref{alg:AD_shape} and Fig.~\ref{fig_decomposition_based_on_shape}. In practice, we set $T_{shape}$ as 0.9. 
Table \ref{gland_results_1} demonstrates the usefulness of shape based AD when segmentation problems contain objects with several shape types.

%
\vspace{-0.1in}
\begin{algorithm}
  \caption{Object-shape based annotation decomposition}
  \label{alg:AD_shape}
\footnotesize
  \begin{algorithmic}[1]
    \Function{AnnotationDecomposition2}{$y_i \in \{1,2, \ldots, K\}^{m\times n}$, $T_{shape}$}
            \State{$y_i^{convex}$ $\gets$ a new array of size $m\times n$ with all 0;}
            \State{$y_i^{concave}$ $\gets$ a new array of size $m\times n$ with all 0;}
            \For{every object $p$ in $y_i$}
                \State Compute the convex hull $p^{convex}$ of $p$
                \State $ratio$ = size($p$)/size($p^{convex}$)
                \If{$ratio > {T}_{shape}$ ($p$ is of a convex-like shape)}
                    \State{Add object $p$ to $y_i^{convex}$}
                \Else
                    \State{Add object $p$ to $y_i^{concave}$}
                \EndIf
            \EndFor
        \State \Return{$y_i^{convex}$ and $y_i^{concave}$}
    \EndFunction
  \end{algorithmic}
\end{algorithm}
\vspace{-0.3in}

\subsubsection{Based on image-level information.}
Image-level information, statistics, and cues can be utilized for annotation map decomposition. For example, if images contain one or multiple foreground objects, we can decompose the segmentation maps based on the number of objects appeared in an image. As the number of objects could only be revealed at a global level or deeper layer in a deep learning model, this decomposition method pushes a learning model to be more aware of global and higher-level information when generating segmentation results. An exact annotation decomposition procedure based on the image-level number of objects is given in Algorithm \ref{alg:AD_imagelevel}. 
In Table \ref{lymph_node_results}, we show the effectiveness of image-level information based AD for lymph node segmentation in ultrasound images.


\vspace{-0.1in}
\begin{algorithm}
  \caption{Image-level information based annotation decomposition}
  \label{alg:AD_imagelevel}
\footnotesize
  \begin{algorithmic}[1]
    \Function{AnnotationDecomposition3}{$y_i \in \{1,2, \ldots, K\}^{m\times n}$}
            \State{$y_i^{single\_obj}$ $\gets$ a new array of size $m\times n$ with all 0;}
            \State{$y_i^{multiple\_obj}$ $\gets$ a new array of size $m\times n$ with all 0;}
            \If{$y_i$ contains only one object} 
                \State{$y_i^{single\_obj} \gets y_i$}
            \EndIf
            \If{$y_i$ contains multiple objects} 
                \State{$y_i^{multiple\_obj} \gets y_i$}
            \EndIf
        \State \Return{$y_i^{single\_obj}$ and $y_i^{multiple\_obj}$}
    \EndFunction
  \end{algorithmic}
\end{algorithm}
\vspace{-0.4in}

\begin{figure}[t]
  \centering
  \includegraphics[width=4.0in]{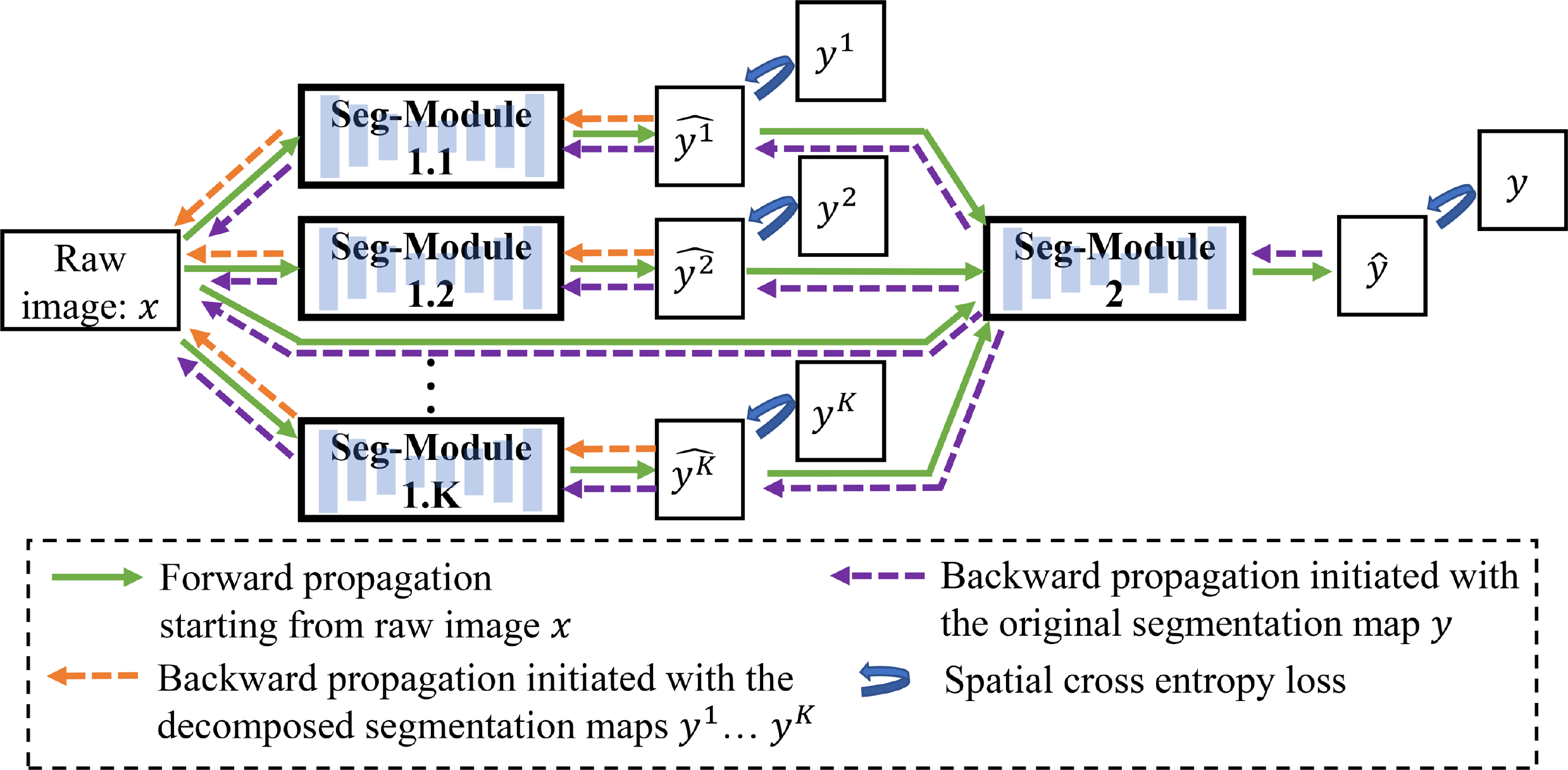}
  \caption{The $K$-to-1 deep network framework for our new decompose-and-integrate learning scheme. $y^{k}$, $k=1,2, \ldots, K$, are decomposed segmentation annotation maps, and $y$ is the original segmentation annotation map.}
\label{fig_K-1}
\end{figure}

\subsection{The $K$-to-1 deep network for decompose-and-integrate learning}\label{sec_K1}
 Suppose every original annotation map $y_i$, $i = 1,2, \ldots, h$, is decomposed into $K$ annotation maps $y^k_i$, $i = 1,2, \ldots, h$ and $k = 1,2, \ldots, K$. We aim to model each sub-segmentation problem using a deep learning segmentation module with its own set of parameters. Then another modeling procedure is applied on top of these $K$ modules to form the final solution of the original segmentation problem.

 Thus, we propose a new $K$-to-1 deep network framework for implementing our above decompose-and-integrate learning scheme. Fig.~\ref{fig_K-1} shows an overview of our $K$-to-1 deep network. The modules (e.g., Seg-Module 1.1, Seg-Module 2) used in this 
 network can be changed according to the type of images (e.g., 2D or 3D images) of the specific segmentation problem. The full model can be trained in end-to-end manner. Let the function of the overall $K$-to-1 network be denoted as $f_{complete}$, and the function of Seg-Module $1.k$ be denoted as $f_{1.k}$. The overall loss for the decompose-and-integrate learning scheme is defined as:

\begin{equation}\label{eq5} 
\frac{1}{h}\sum_{i=1}^{h}(\mathcal{L}(f_{complete}(x_i),y_i) + \lambda \sum_{k=1}^{K}\mathcal{L}(f_{1.k}(x_i),y^k_i))
\end{equation}
\noindent

where $\mathcal{L}$ is the spatial cross entropy loss, and $\lambda$ is set as simple as a normalization term $\frac{1}{K}$. We aim to minimize the above function with respect to the parameters of $f_{complete}$ and $f_{1.k}$ for $k = 1,2, \dots, K$. 





\section{Experiments and Results}\label{sec:exp}
We conduct experiments on three datasets. The 3D cardiovascular segmentation dataset \cite{pace2015interactive} contains two classes of foreground objects (myocardium and great vessels), which have close spatial relations. Thus, we apply object-class based annotation decomposition (AD) to this dataset. The gland segmentation dataset \cite{sirinukunwattana2017gland} contains glands that have quite different shapes (from concave shape to convex shape); hence shape convexity based annotation decomposition (AD) is applied to this dataset. Our in-house lymph node dataset 
contains the lymph node areas of 237 patients in ultrasound images (one image may contain one or more lymph nodes). Thus, image-level information based AD is applied to this dataset.

\begin{table}[t]
\centering
\caption{Comparison of segmentation results on the HVSMR dataset.}
\label{tabHVSMR1}
\scriptsize
\resizebox{\columnwidth}{!}{
\begin{tabular}{|c ccc ccc c|}
    \hline
    \multirow{2}{*}{ Method } &   \multicolumn{3}{c}{Myocardium} & \multicolumn{3}{c}{Blood pool}  &  \multirow{2}{*}{\shortstack{Overall \\ score}} \\  \cline{2-7}
    & Dice   &  ADB     &  Hausdorff   & Dice   &  ADB    &  Hausdorff  &  \\ 
    \hline
    3D U-Net \cite{cciccek20163d} 
    &  $0.694$  &  $1.461$  & $10.221$  &   $ 0.926$ & $ 0.940 $  &  $ 8.628$   &  $ -0.419 $ \\ 
    \hline 
    VoxResNet \cite{chen2017voxresnet}  & $ 0.774 $   &  $ 1.026  $  &  $ 6.572  $  &  $ 0.929  $ & $ 0.981 $  & $ 9.966 $     & $ -0.202 $  \\ 
    \hline 
    DenseVoxNet  \cite{yu2017automatic} &  $ 0.821  $  &  $ 0.964  $  &  $ 7.294  $  & $ 0.931  $  &  $ 0.938  $  &  $ 9.533  $    &  $ -0.161 $ \\ 
    \hline
     Ensemble Meta-learner \cite{zheng2018new}& 0.823  &   0.685    &  3.224   &  0.935  &   0.763   &  5.804  &  0.215  \\ %
     \hline
   Class-AD + $K$-to-1 DenseVoxNet ({\bf ours})&  $ 	0.839  $  &  $ 0.744  $  &  $ 3.500  $  & $ 0.941  $  &  $ 0.658  $  &  $  5.973  $    &  $ {\bf 0.223} $ \\ 
    \hline
{Ablation Study}: &  &  &  &  &  & &  \\
    \hline

     Large DenseVoxNet &  $ 0.804 $  &  $ 0.847  $  &  $ 3.980  $  & $ 	0.935  $  &  $ 	0.756  $  &  $  7.706  $    &  $ {0.079} $ \\ 
      \hline
          2-stacked DenseVoxNet&  $ 0.837  $  &  $ 0.797  $  &  $ 3.405 $  & $ 0.939  $  &  $ 0.629  $  &  $  7.529  $    &  $ {0.167} $ \\ 
    \hline
    $K$-to-1 DenseVoxNet w/o AD &  $ 0.824  $  &  $ 0.776  $  &  $ 3.619  $  & $ 0.940  $  &  $ 0.677  $  &  $  6.632  $    &  $ {0.177} $ \\ 
    \hline
\end{tabular}
    }
\end{table}

{\bf Implementation details.}
The input window size of the deep learning segmentation models we use is set as $64\times64\times64$ for 3D experiments and $192\times192$ for 2D experiments. During training, random cropping, rotation, and flipping are applied. Since the images in each dataset are larger than the 
model window size, there are virtually many more 
samples for model training than the number of images in each dataset. The Adam optimizer is used for model training. The mini-batch size is set as 8. The maximum number of training iteration is set to 60000. We find that usually 60000 iterations using Adam are sufficient for an FCN-type model to converge for a moderate sized training set. The learning rate is set as 0.0005 initially, and decreased to 0.00005 after 30000 iterations. 

{\bf 3D cardiovascular segmentation in MR images.}
The HVSMR dataset \cite{pace2015interactive} seeks to segment myocardium and great vessels (blood pool) in 3D cardiovascular MR images. The ground truth of the test data is not available to the public; the evaluations are done by submitting segmentation results to the organizers' server. We experiment with the object class based AD for this dataset. Table \ref{tabHVSMR1} shows that our AD combined with $K$-to-1 network (utilizing DenseVoxNets) achieves state-of-the-art performance on this dataset. In the ablation study part of Table \ref{tabHVSMR1}, we compare our full model with K-to-1 network without AD (where $y_i^k=y_i, i = 1,2, \ldots, h$, and $k = 1,2, \ldots, K$), a 2-stacked DenseVoxNets, and a large-size DenseVoxNet that uses a similar amount of parameters as the $K$-to-1 DenseVoxNets. The ablation study results confirm the effectiveness of our decompose-and-integrate learning scheme.

{\bf Gland segmentation in H\&E stained images.} This dataset \cite{sirinukunwattana2017gland} contains 85 training images (37 benign (BN), 48 malignant (MT)), 60 testing images (33 BN, 27 MT) in part A, and 20 testing images (4 BN, 16 MT) in part B. We modify the original CUMedNet \cite{chen2016deep} to make it deeper with two more encoding and decoding blocks (denoted as CUMedNet$^+$). We run all the experiments for the $K$-to-1 network and ablation study 5 times. Table \ref{gland_results_1} shows the mean performance and standard derivations. Compared with the state-of-the-art models, our AD + $K$-to-1 network (utilizing CUMedNet$^+$) yields considerably better segmentation results. In ablation study (the bottom part of Table \ref{gland_results_1}), we compare AD + $K$-to-1 network with $K$-to-1 network without AD, a 2-stacked CUMedNet$^+$, and a large-size CUMedNet$^+$.



\begin{table*}[t]
\small
\centering
	\caption{Comparison of segmentation results on the gland segmentation dataset.}
	\label{gland_results_1}
	\scriptsize
\resizebox{\columnwidth}{!}{
\begin{tabular}{|ccccccc|}

  \hline
  \multirow{2}{*}{Method}&\multicolumn{2}{c }{$F_1$ Score}&\multicolumn{2}{c }{ObjectDice}&\multicolumn{2}{c| }{ObjectHausdorff}\\\cline{2-7}
  &\multicolumn{1}{c }{part A}&\multicolumn{1}{c }{part B}&\multicolumn{1}{c }{part A}&\multicolumn{1}{c }{part B}&\multicolumn{1}{c }{part A}&\multicolumn{1}{c|}{part B}\\
    \hline
    CUMedVision \cite{chen2016dcan}& 0.912 & 0.716 & 0.897  &0.718  &45.418  &160.347 \\
  \hline

  Multichannel2 \cite{xu2016gland}& 0.893  & 0.843  & 0.908  &0.833&{44.129}&116.821 \\
   \hline
   MILD-Net \cite{graham2018mild} &0.914 & {0.844} & \textbf{0.913} &0.836&{41.54}&105.89  \\
  \hline
CUMedNet \cite{chen2016deep}$^+$ & 0.907$\pm$0.007  & 0.835$\pm$0.009   & 0.893$\pm$0.007  &0.832$\pm$0.008 & 49.97$\pm$2.12 &113.40$\pm$7.22 \\

    \hline
  Shape-AD + $K$-to-1 {\bf(ours)} & \textbf{0.923}$\pm$0.002  & \textbf{0.861}$\pm$0.004   & 0.910$\pm$0.004  &\textbf{0.846}$\pm$0.001 & \textbf{40.79}$\pm$1.72 & \textbf{101.42}$\pm$1.49 \\
    \hline
{Ablation Study}: &  &  &  &  &  &  \\
    \hline
    Large CUMedNet$^+$ &0.918$\pm$0.005  & 0.817$\pm$0.021   & 0.903$\pm$0.002  &0.827$\pm$0.012 & 43.81$\pm$1.39 &{109.43$\pm$5.39} \\

  \hline
  
  2-stacked CUMedNet$^+$ & 0.914$\pm$0.002  & 0.830$\pm$0.009   & 0.908$\pm$0.001  &0.844$\pm$0.002 & 45.32$\pm$1.05 &{101.43$\pm$2.25} \\
    \hline
    $K$-to-1 w/o AD & 0.915$\pm$0.007  & 0.829$\pm$0.008   & 0.898$\pm$0.007  &0.831$\pm$0.004 & 45.23$\pm$3.71 &{108.92$\pm$4.74} \\

  \hline

\end{tabular}
}
\end{table*}

{\bf Lymph node segmentation in ultrasound images.} We collected 
patients' lymph node ultrasound images. We use 137 images for model training, and  
100 images for model testing. The image size is $1080 \times 768$. There is no identity overlap between the training data and testing data. The AD procedure follows Algorithm \ref{alg:AD_imagelevel}. Table \ref{lymph_node_results} demonstrates that AD + $K$-to-1 network can effectively improve lymph node segmentation performance in ultrasound images. 

\begin{table*}[t]
\centering
	\caption{Comparison of segmentation results on the lymph node segmentation dataset.}
	\label{lymph_node_results}
\scriptsize

\begin{tabular}{|ccccc|}
  \hline
  {Method}&{IoU}&{Precision}&{Recall}&{$F_1$ Score}\\
    \hline
U-Net \cite{ronneberger2015u}& {0.661} & {0.834} & {0.7607}& {0.7957}\\\hline
Deeper U-Net  & {0.7369} & {0.8555} & {0.8416} & {0.8485}\\\hline
CUMedNet \cite{chen2016deep}$^+$& {0.7595} & {0.8472} & {0.8801
} & {0.8633} \\
    \hline
Image-level-AD + $K$-to-1 ({\bf ours})& \bf{0.8102} & \bf{0.9012} & \bf{0.8893} &\bf{0.8952}\\

    \hline
Ablation Study: &  &  &  &\\
    \hline
        Large CUMedNet$^+$& {0.7795} & {0.8808} & { 0.8714} & {0.8761} \\
        \hline
    2-stacked CUMedNet$^+$& {0.7759} & {0.876} & { 0.8716} & {0.8738} \\
        \hline
 $K$-to-1 w/o AD & {0.7842} & {0.8798} & {0.8783} & {0.8790} \\
    \hline

\end{tabular}
\end{table*}

\section{Conclusions}
In this paper, we developed a new decompose-and-integrate learning scheme for medical image segmentation. Our new learning scheme is well motivated, sound, and quite flexible. Comprehensive experiments on multiple datasets show that our new learning scheme is effective in improving segmentation performance.

\bibliographystyle{splncs03}
\bibliography{sample}
\end{document}